\documentstyle[12pt]{article}
\begin{document}

\begin{center}
IMPORTANCE OF BARYON-BARYON COUPLING IN HYPERNUCLEI\\
\vspace*{36pt}
B.\ F.\ Gibson\\
\vspace{3pt}
Theoretical Division, Los Alamos National Laboratory\\
Los Alamos, New Mexico  87545, U.~S.~A.\\
\vspace{2pt}
and\\
\vspace{2pt}
Institut f\"ur Kernphysik, Forschungszentrum\\
52425 J\"ulich, Germany\\
\vspace{12pt}
I.\ R.\ Afnan\\
\vspace{3pt}
School of Physical Sciences, The Flinders University of South Australia\\
Adelaide, South Australia 5001, Australia\\
\vspace{12pt}
J.\ A.\ Carlson\\
\vspace{3pt}
Theoretical Division, Los Alamos National Laboratory\\
Los Alamos, New Mexico  87545, U.~S.~A.\\
\vspace{12pt}
D.\ R.\ Lehman\\
Department of Physics, The George Washington University\\
Washington, District of Columbia 20052, U.~S.~A.
\vspace*{24pt}

ABSTRACT\\
\vspace{8pt}
\parbox [t] {5in}{{\small The $\Lambda N - \Sigma N$ coupling in
$\Lambda$--hypernuclei and $\Lambda \Lambda - \Xi N$ coupling in
$\Lambda \Lambda$--hypernuclei produce novel physics not observed
in the conventional, nonstrange sector.  Effects of
$\Lambda \leftrightarrow \Sigma$ conversion in $^3_{\Lambda}$H are
reviewed.  The role of $\Lambda N - \Sigma N$ coupling suppression
in the $A=4,5$ $\Lambda$--hypernuclei due to Pauli blocking is
highlighted, and the implications for the structure of
$^{10}_{\;\, \Lambda}$B are explored.  Suppression of
$\Lambda \Lambda - \Xi N$ conversion in
$^{\;\;\, 6}_{\Lambda \Lambda}$He is hypothesized
as the reason that the $< V_{\Lambda \Lambda} >$ matrix element
is small.  Measurement of $^{\;\;\, 4}_{\Lambda \Lambda}$H is proposed to
investigate the full $\Lambda \Lambda - \Xi N$ interaction.
The implication for $\Lambda \Lambda$ analog states is discussed.}}
\end{center}

\pagebreak
\vspace{16pt}
\noindent {\bf 1. Introduction}

\vspace{4pt}
Since the discovery of the first hyperfragment in a balloon flown
emulsion stack some forty years ago, physicists have labored to
understand how the addition of the strangeness degree of freedom
($S = -1$ for $\Lambda$ and $\Sigma$ and $S=-2$ for $\Xi$ and
$\Lambda \Lambda$) alters our picture of nuclei and the
baryon-baryon force.  Because the $\Lambda$ and $\Sigma$ masses
differ markedly from the nucleon mass, $SU(3)$
symmetry is broken.  How it is broken is a question of importance
to our fundamental understanding of the baryon-baryon interaction.
Furthermore, the inclusion of an $S\neq0$ degree of freedom
(flavor) in the nucleus adds a third dimension to our evolving
picture of nuclei.  We have discovered that the physics of
hypernuclei is unique --- different from the conventional
nuclear physics that we have investigated for more than
sixty years.  Unlike the $NN$ interaction which is dominated
by one pion exchange (OPE), the $\Lambda N$ interaction has
no OPE component.  Thus, the shorter range aspects of the
nuclear force may no longer be hidden under
the long range OPE mechanism.  $\Lambda N - \Sigma N$ coupling
plays a significant role in the hyperon-nucleon ($YN$) force
in contrast to the very limited role that $NN-N \Delta$
coupling appears to play in the $S=0$ sector.  This is due
in part to the fact that
$m_{\Sigma} - m_{\Lambda} \simeq 80$ MeV,
whereas $m_{\Delta} - m_N \simeq 300$ MeV; furthermore,
the $\Delta$ is a broad resonance that decays via the strong
interaction, so that its influence may be spread over a
broad energy range.  However,
it is also possible that baryon coupling within the octet
differs from the coupling between members of the baryon octet
and the isobar decuplet.  Extending the analysis to the $S=-2$
sector of the octet, one would anticipate strong
$\Lambda \Lambda - \Xi N$ ($YY$) coupling;
$m_{\Xi N} - m_{\Lambda \Lambda} \simeq 25$ MeV.
Pauli blocking of the nucleon in the
coupled-channel interaction may well explain why
one extracts a relatively weak matrix element from the three
observed $\Lambda \Lambda$--hypernuclei, in stark contrast
to the strong attraction that one expects because
it falls in the same multiplet as the $nn$ interaction.

We will review here some of the interesting aspects of
hypernuclei which demonstrate that $S \neq 0$ physics
is different.  In addition, we will explore how few-baryon
systems can be used to constrain our models of the $YN$ and
$YY$ interactions,  and we will reflect upon what we have so far
achieved in understanding the baryon-baryon force in this sector.

\vspace{16pt}
\noindent {\bf 2. The Role of $\Lambda N - \Sigma N$ Coupling}

\vspace{4pt}
We have only to examine the few-body data to see that
$\Lambda N - \Sigma N$ coupling plays a significant role in
$\Lambda$-hypernuclei.  In Table 1 we summarize the $\Lambda$
separation energies,
$$
B_{\Lambda} (_{\Lambda}A) = B (_{\Lambda}A) - B (A - 1) \; ,
$$
for the $s$-shell $\Lambda$-hypernuclei {\cite{jur73,dav67,dav91}}
along with $\gamma$ de-excitation energies {\cite{bam73,bej79}},
where they have been measured.

\begin{table}[bht]
\begin{flushleft}
{\bf Table 1.}  Hypernuclear $\Lambda$-separation energies and
 excitation energies in MeV
\end{flushleft}
\begin{tabular}{ccccccc} \hline
\hspace*{1cm} & hypernucleus & \hspace{2.5cm} & B$_{\Lambda}$ &
  \hspace{1.5cm} & E$_{\gamma}$ & \hspace{1cm} \\ \hline
\\
\hspace*{1cm} &  $^3_{\Lambda}$H & \hspace{2.5cm} & 0.13$\pm$.05 &
  \hspace{1.5cm} & \hspace{1cm} \\
\\
\hspace*{1cm} &  $^4_{\Lambda}$H & \hspace{2.5cm} & 2.04$\pm$.04 &
  \hspace{1.5cm} & 1.04$\pm$.04 & \hspace{1cm} \\
\\
\hspace*{1cm} & $^4_{\Lambda}$He & \hspace{2.5cm} & 2.39$\pm$.03 &
  \hspace{1.5cm} & 1.15$\pm$.04 & \hspace{1cm} \\
\\
\hspace*{1cm} & $^5_{\Lambda}$He & \hspace{2.5cm} & 3.10$\pm$.02 &
  \hspace{1.5cm} & \\
\\ \hline
\end{tabular}
\end{table}

\vspace{6pt}

The charge symmetry breaking (CSB) in the $A=4$ isodoublet is
obvious:   $^4_{\Lambda}$He is more bound than $^4_{\Lambda}$H
by $\Delta B_{\Lambda} =$ 0.35 MeV.   The $\Lambda p$
interaction is stronger than the $\Lambda n$ interaction.  [This
CSB is almost 3 times that estimated for the $^3$H -- $^3$He
system once the Coulomb energy is taken into account:
$B(^3 {\rm H}) - B(^3 {\rm He}) - E_C \simeq 8.48 - 7.72 - 0.64
\simeq 0.12$ MeV].
One needs no theoretical calculation to see the CSB in the
$\Lambda$-hypernuclei separation energies, because the Coulomb
energy of the $^3$He core essentially cancels that of
$^4_{\Lambda}$He.  The second order Coulomb effect due to the
compression of the $^3$He core in $^4_{\Lambda}$He by the addition
of the $\Lambda$ actually increases $\Delta B_{\Lambda}$ by
0.01 -- 0.02 MeV.   This CSB provides a clear signal for the
importance of $\Lambda N - \Sigma N$ coupling in the $YN$
interaction.  The ($\Sigma^+, \Sigma^0, \Sigma^-$) mass splitting
is some 9 MeV, or about 10\% of the $\Lambda - \Sigma$ mass
difference.  Including this alone accounts for a large part of
the CSB in the YN interaction.

That $\Lambda N - \Sigma N$ coupling is more significant in
$\Lambda$-hypernuclei than is $NN - N\Delta$ coupling in
conventional nuclei can be seen by comparing $\Lambda$
separation energies from $^3_{\Lambda}$H, $^4_{\Lambda}$He,
and $^5_{\Lambda}$He with neutron separation energies from
$^2$H, $^3$H, and $^4$He.   In the nonstrange sector we know
that the ratio of neutron separation energies for neighboring
s-shell nuclei is close to a constant:
$$
B_n(^3{\rm H})/ B_n(^2{\rm H}) \simeq \frac{6}{2} = 3
$$
and
$$
B_n(^4{\rm He})/ B_n(^3{\rm He}) \simeq \frac{20}{6}
\simeq 3 \: .
$$
\noindent Thus, one might anticipate that
\begin{eqnarray*}
B_{\Lambda}(^5_{\Lambda}{\rm He}) &\simeq& 3 \;
B_{\Lambda}(^4_{\Lambda}{\rm He})\\
&\simeq& 6 \; {\rm MeV}
\end{eqnarray*}
for model calculations in which $V_{\Lambda N}$ is fitted
to the low energy $\Lambda N$ scattering parameters and
reproduces the observed value of
$B_{\Lambda}(^4_{\Lambda}{\rm He}) \simeq 2$ MeV.  This is,
in fact, confirmed by several model calculations
{\cite{bod68,her67,gib72,gal75}} using central potentials to
represent the $\Lambda N$ force.  In contrast, the
experimental value for $B_{\Lambda}(^5_{\Lambda}$He)
is $3.1$ MeV {\cite{jur73,dav67,dav91}}.
Similarly, for the hypertriton, one might estimate
\begin{eqnarray*}
B_{\Lambda}(^3_{\Lambda}{\rm H}) &\simeq& 1/3 \;
B_{\Lambda}(^4_{\Lambda}{\rm He})\\
&\simeq& 0.7 \; {\rm MeV} \: .
\end{eqnarray*}
This is again confirmed by several model calculations
using central potentials to represent the $\Lambda N$
interaction {\cite{dab73,bfg75,ira90}}.  In contrast, the
experimental value for $B_{\Lambda}(^3_{\Lambda}$H) is
$0.13$ MeV.

In other words, our carefully developed intuition coming from
the study of conventional, nonstrange nuclei fails to
provide a sound means of extrapolating into the
strangeness sector.  The physics of $\Lambda$--hypernuclei
is new and different.  A large part of this difference
can be traced to $\Lambda N - \Sigma N$ coupling.
(If $NN - N\Delta$ coupling played as significant a role in
the nonstrange sector, then our modeling of nuclei would
have required more sophistication.   Our shell model
calculations, which assume a mean field interaction
independent of the quantum numbers of the core, would have
required serious modification.)  Why do we suggest that a
major factor in the anomalously small binding of
$^5_{\Lambda}$He is the strong suppression of
$\Lambda N - \Sigma N$ coupling?
Converting a T=0 $\Lambda$ into a T=1
$\Sigma$ requires the simultaneous conversion of the $^4$He
core of $^5_{\Lambda}$He into a T=1, even-parity $^4$He$^*$
excited state {\cite{bfg69}}.  Such excited states lie high
in the continuum.  They also have wave functions with more
structure than the $^4$He ground state, which reduces the
transition matrix element.   Thus, the strength of the full
coupled-channel
\begin{eqnarray*}
\left( \begin{array}{c}
V_{\Lambda N} \hspace{0.25in} V_{XN} \\
V_{XN}  \hspace{0.25in} V_{\Sigma N} \\
\end{array} \right)
\end{eqnarray*}
interaction is expected to be significantly reduced in
$^5_{\Lambda}$He, as the off-diagonal transitions are
partially Pauli blocked.  That $\Lambda N - \Sigma N$
coupling is suppressed in $^5_{\Lambda}$He is confirmed by
Monte Carlo variational calculations {\cite{JAC91}} using the
contemporary Nijmegen soft core potential {\cite{nijSC}},
although the $\Lambda N - \Sigma N$ coupling is so strong
in this Nijmegen model that the $A=5$ system is actually
unbound in the calculation.  (The anomalously small binding
has also been interpreted as a repulsive $\Lambda NN$
three-body force in models which neglect explicit
$\Lambda N - \Sigma N$ coupling {\cite{bod88}}.)

Similarly, using a $YN$ interaction that incorporates
$\Lambda N - \Sigma N$ coupling leads to a reduction in the binding of
$^3_{\Lambda}$H compared to that from a single-channel $\Lambda N$
potential {\cite{ira90}}.  The hypertriton can be thought of
principally as a $\Lambda$ bound tenuously to a deuteron.  The
$\Sigma$ isospin ($T=1$) differs from that of the $\Lambda$ ($T=0$).
Therefore, converting the $\Lambda$ into a $\Sigma$ requires
simultaneously converting the deuteron core into a $^1$S$_0$ $NN$
pair.   The $NN$ $^1$S$_0$ interaction is much weaker than the
tensor dominated $^3$S$_1$--$^3$D$_1$ force that binds the
deuteron.  Thus, incorporating $\Lambda N - \Sigma N$ coupling
into the interaction model leads to a reduction in the
$^3_{\Lambda}$H binding.  Such an effect might be interpreted
as a repulsive $\Lambda NN$ three-body potential.  However,
if one formally eliminates the $\Sigma N$ channel (producing
an energy-dependent $\Lambda N$ potential plus a true
$\Lambda NN$ potential), one finds in explicit calculations that
the true $\Lambda NN$ three-body force is attractive {\cite{ira90}}.
Nonetheless, it is the case that the anomalously small binding
of the hypertriton is the result of strong $\Lambda N - \Sigma N$
coupling in the $S=-1$ baryon-baryon force.

\vspace{16pt}
\noindent {\bf 3. The $YN$ Scattering Data}

\vspace{4pt}
The physics of the $S = -1$ sector clearly differs from that
of the $S = 0$ sector.  Can we understand the $\Lambda$
separation energies in terms
of the $S = -1$ hyperon--nucleon  interaction?  Unfortunately
the $\Lambda N$, $\Sigma N$ data are very limited.  There
are some 600 events{\cite{alx64,eng66,sec68}} below 300
MeV/c and another 250 events{\cite{kad71}} between 300 and
1500 MeV/c --- total cross sections and some angular data.
The lack of a significant data base severely inhibits our
ability to provide a definitive analysis of hypernuclear
data.

The Nijmegen group has taken the lead in modeling
{\cite{nijD,nijF,nijSC}} the $YN$ potential by fitting a
one-boson-exchange (OBE) hypothesis to the combined $NN$
and $YN$ data base using $SU(3)$ constraints.  Their
pioneering effort to construct such potentials has been
summarized by deSwart in a recent Seminar on
the hyperon-nucleon interaction {\cite{jjs94}}.  Timmermans
elaborated further upon the model {\cite{rt94}}.  The J\"ulich
group have also produced $YN$ potential models
{\cite{jul89,hol92,jul92}}, both energy dependent and of
the OBE form (in q-space).  Reuber discussed the J\"ulich
approach at the same meeting {\cite{agr94}}.  The models are
not the same.  For example the F/D ratios differ significantly,
which affects the strength in the $\Sigma N$ channel.

\begin{table}[bht]
\begin{flushleft}
{\bf Table 2.}  The charge-independent scattering lengths and
effective ranges in fm for the $YN$ potential models listed
\end{flushleft}
\begin{tabular}{clcc|ccccccccc} \hline
\hspace*{1cm}& Model & Ref. &\hspace{1cm}&\hspace{2cm}& a$^s$ &
  \hspace{.5cm}& r$^s_o$ &\hspace{2cm}& a$^t$ &\hspace{.5cm}& r$^t_o $
  &\hspace{1cm} \\ \hline
 & & & &  & & & & & & & & \\
\hspace*{1cm}& Nijmegen D & {\cite{nijD}} &\hspace{1cm}&\hspace{2cm}&
 -1.90 &\hspace{.5cm}& 3.72 &\hspace{2cm}& -1.96 &\hspace{.5cm}& 3.24
  &\hspace{1cm} \\
 & & & &  & & & & & & & & \\
\hspace*{1cm}& Nijmegen F & {\cite{nijF}} &\hspace{1cm}&\hspace{2cm}&
 -2.29 &\hspace{,5cm}& 3.17 &\hspace{2cm}& -1.88 &\hspace{.5cm}& 3.36
  &\hspace{1cm} \\
 & & & &  & & & & & & & & \\
\hspace*{1cm}& Nijmegen SC& {\cite{nijSC}}&\hspace{1cm}&\hspace{2cm}&
 -2.78 &\hspace{.5cm}& 2.88 &\hspace{2cm}& -1.41 &\hspace{.5cm}& 3.11
  &\hspace{1cm} \\
 & & & &  & & & & & & & & \\
\hspace*{1cm}& J\"ulich A & {\cite{jul92}}&\hspace{1cm}&\hspace{2cm}&
 -1.56 &\hspace{.5cm}& 1.43 &\hspace{2cm}& -1.59 &\hspace{.5cm}& 3.16
  &\hspace{1cm} \\
 & & & &  & & & & & & & & \\
\hspace*{1cm}& J\"ulich \~A&{\cite{jul92}}&\hspace{1cm}&\hspace{2cm}&
 -2.04 &\hspace{.5cm}& 0.64 &\hspace{2cm}& -1.33 &\hspace{.5cm}& 3.91
  &\hspace{1cm} \\
 & & & &  & & & & & & & & \\
 \hline
\end{tabular}
\end{table}

\vspace{6pt}

In Table 2 we summarize for several of the models their
charge-independent low-energy scattering parameters  --- the
scattering lengths and effective ranges.  A quick comparison
illustrates the need for better $YN$ data to constrain
the potential parameters.  Nonetheless, based upon the
models, the following observations can be made.
The $\Lambda N - \Sigma N$ coupling is strong.   The
$\Lambda (T=0) \times N(T = \frac{1}{2})$
direct interaction contains no OPE, although a second
order OPE force is generated through
$\Lambda N - \Sigma N$ coupling.   The short range
aspects of the baryon-baryon force, such as heavy meson
exchange, play a more important role.  Oka recently
reviewed the Tokyo quark-cluster modeling of the short
range properties of the $YN$ interaction {\cite{oka94}}.
It is in the YN sector that one begins to find differences
between the OBE and quark cluster models of the
baryon-baryon force.  For example, the OBE potential
exhibits strong short range attraction in some $\Sigma N$
channels, whereas the quark cluster approach appears to
yield consistently short range repulsion.

\vspace{16pt}
\noindent {\bf 4. The $A=4$ Isodoublet}

\vspace{4pt}

\vspace{8pt}
\noindent {\it 4.1. Charge Symmetry Breaking}
\vspace{4pt}

We have remarked already that the large CSB in the $A=4$
isodoublet ground-state binding energies provides strong
evidence for the important role of $\Lambda N - \Sigma N$
coupling in the $YN$ interaction.  How well can this CSB
be understood in terms of the free $\Lambda N$ interaction?
The Nijmegen models predict a noticeable charge symmetry
breaking:  $V_{\Lambda p} \neq V_{\Lambda n}$. [Recall
that strong $\Lambda - \Sigma$ conversion leads to CSB
because of the mass difference in the $\Sigma$ isospin
triplet.]  The $\Lambda p$ and $\Lambda n$ scattering
lengths and effective ranges from Nijmegen model D are
compiled in Table 3.

\begin{table}[bth]
\begin{flushleft}
{\bf Table 3.}  The charge-dependent scattering lengths and
effective ranges in fm for the Nijmegen model D
\end{flushleft}
\begin{tabular}{ccc|ccccccccc} \hline
\hspace*{1cm}& channel    &\hspace*{1cm}& \hspace*{1cm}& a$^s$ &
  \hspace{1cm}& r$^s_0$ &\hspace{2cm}& a$^t$ &\hspace{1cm}& r$^t_0 $ &
  \hspace{1cm}  \\ \hline
 &  &  &   & & & & & & &  \\
\hspace*{1cm}& $\Lambda p$ &\hspace*{1cm}&\hspace*{1cm}& -1.77 &
  \hspace{1cm}& 3.78 &\hspace{2cm}& -2.06 &\hspace{1cm}& 3.18 &
  \hspace{1cm} \\
 &  &  &   & & & & & & &  \\
\hspace*{1cm}& $\Lambda n$ &\hspace*{1cm}&\hspace*{1cm}& -2.03 &
  \hspace{1cm}&   3.66 &\hspace{2cm}& -1.84 &\hspace{1cm}& 3.32 &
  \hspace{1cm} \\
 &  &  &   & & & & & & &  \\  \hline
\end{tabular}
\end{table}

\vspace{6pt}

One can use a folding prescription to generate an effective
two-body ($\Lambda$-nucleus) potential with which to calculate
$B_{\Lambda} (^4_{\Lambda}$He) and $B_{\Lambda}(^4_{\Lambda}$H).
The resulting $\Delta B_{\Lambda}$ from this approach (in
terms of two-body dynamics) using s-wave separable potentials
fitted to the low-energy scattering parameters in Table 3 is
$0.21$ MeV.  If one allows for compression of the nuclear core
(due to the added binding of the $\Lambda$), one can increase
this to 0.24 MeV for a 5\% core compression.  In contrast, using
exact four-body equations and the identical separable
potentials, one finds a value of $\Delta B_{\Lambda} = 0.43$
MeV {\cite{GL79}}.  That is, true 4-body dynamics yields a value
for $\Delta B_{\Lambda}$ about twice that which comes from a
model utilizing the same potentials but approximate two-body
dynamics.

This perhaps surprising result can be simply understood.  One
can demonstrate {\cite{bfg73,GL76}} for simple potentials
that $|a| > |a^{\prime}|$ implies that potential $V$ is
stronger than potential $V^{\prime}$ in an n-body calculation,
where n = 2, 3, 4.  That is, the potential $V$ is uniformly
stronger than $V^{\prime}$, as intuition would lead one to
expect because $a$ represents the overall strength of the
interaction.  In contrast, for $r > r^{\prime}$,
the same relation holds only for $n = 2$.  (The larger the potential
range, the stronger is the potential in a two-body sense.)
However, if $n = 3, 4,$ then $r > r^{\prime}$ means that $V$
is weaker than $V^{\prime}$.   Recall Thomas' result
{\cite{lht35}} from 1935:  If in a three-body system the range
of the force $\rightarrow 0$ for one pair, then the three-body
binding becomes infinite.
Therefore, the CSB in the $A=4$ isodoublet can be understood in
terms of the scattering length and effective range differences
inferred from Table 3.  The spin-singlet interactions tend to
average in the two systems and contribute little to the $\Lambda$
separation energy difference; on the other hand, the spin-triplet
interaction in $^4_{\Lambda}$He is pure $\Lambda p$ while that in
$^4_{\Lambda}$H is pure $\Lambda n$.  We see that $\Delta a$ and
$\Delta r$ for the spin-triplet potentials will produce
compensating effects in a two-body, folding potential
approximation (reducing the estimate of CSB in the isodoublet) but
will add constructively in a true four-body calculation.  Thus, we
understand that the difference between the 0.21 MeV and 0.43 MeV
quoted above for identical separable potentials comes from the
dynamics of the calculations.  When one looks at details of
few-body systems, exact model calculations can play a crucial role
in a correct interpretation.

We note in closing this discussion that tensor forces cannot be
neglected; they are known to be less effective in binding few-body
systems than are central forces.  When a tensor component is
included in the the $\Lambda N$ spin-triplet interaction, the
binding energy difference in the $A=4$ isodoublet is reduced from
0.43 MeV to 0.37 MeV.


\vspace{8pt}
\noindent {\it 4.2. The Excited States}
\vspace{4pt}

The $^4_{\Lambda}$H/$^4_{\Lambda}$He system is even more
interesting because there exists a particle-stable
excited state for each species.  The ground states
are $0^+$, the excited states are $1^+$.  It is therefore
tempting to argue that the measured $E_{\gamma}$ transition
energies provide a direct determination of the $\Lambda N$
spin-spin interaction.  However, we suggest 1) that is not
the case, 2) that $\Lambda N - \Sigma N$ coupling and the
complex structure of the $A = 4$ $\Lambda$-hypernuclei
must be taken into account in any serious calculation,
and 3) that the improper modeling of $^4_{\Lambda}$H as
$[\Lambda \times ^3$H$]^{[J]}$ is likely the explanation for
the ``missing'' $\gamma$-ray in the $^{10}_{\;\,\Lambda}$B
system.

The existence of particle-stable excited states presents
a unique opportunity to test our models of the $YN$
interaction.  Fitting both the $0^+$ and $1^+$ states
within the same model is a nontrivial task.
$\Lambda N - \Sigma N$ coupling, as a crucial feature of
the $YN$ interaction, plays a large role.  Furthermore, the
complex nature of the four-body system is important.
The $1^+$ state is not just formed from the $0^+$ state
by a spin flip of the $\Lambda$.

\begin{table}[bth]
\begin{flushleft}
{\bf Table 4.}  The $\Lambda N$ scattering lengths and
effective ranges in fm for the Stepien-Rudzka and Wycech
model
\end{flushleft}
\begin{tabular}{ccc|ccccc} \hline
\hspace*{1cm}& channel &\hspace*{2cm}& \hspace*{1cm}& spin-singlet &
  \hspace{3cm}& spin-triplet &\hspace{1cm}  \\ \hline
 &  &  &   & & & &  \\
\hspace*{1cm}& a(fm)   &\hspace*{2cm}&\hspace*{1cm}& -1.97 &
  \hspace{3cm}& -1.95   &\hspace{1cm} \\
 &  &  &   & & & &  \\
\hspace*{1cm}& r$_0$(fm)&\hspace*{2cm}&\hspace*{1cm}&  3.90 &
  \hspace{3cm}&  2.43   &\hspace{1cm} \\
 &  &  &   & & & &  \\ \hline
\end{tabular}
\end{table}
\vspace{6pt}

Consider a model example in which the Stepien-Rudzka and
Wycech {\cite{SRW81}} separable potential approximation
to the Nijmegen model D is employed.  The scattering lengths
and effective ranges are displayed in Table 4.
If one uses effective $\Lambda N$ potentials fitted
to these parameters ({\it i.e.,} explicit $\Lambda N - \Sigma N$
coupling in the $YN$ interaction is approximated by
constructing a $\Lambda N$ potential that reproduces the
scattering length and effective range of the full interaction)
in a full four-body calculation, then one obtains for the total
binding energies {\cite{GL88}}:
$$
B(0^+) = 10.7 \; {\rm MeV}
$$
$$
B(1^+) = 11.7 \; {\rm MeV} .
$$
The order of the states is inverted; the $1^+$ is the ground
state.   The reason is clear if one looks at the average
$\Lambda N$ potential in the two cases.  One finds:
$$
0^+: V_{\Lambda N} = \frac{1}{2} V^S_{\Lambda N} + \frac{1}{2}
V^t_{\Lambda N}
$$
and
$$
1^+: V_{\Lambda N} = \frac{1}{6} V^S_{\Lambda N} + \frac{5}{6}
V^t_{\Lambda N} .
$$
Although $V^S_{\Lambda N}$ from Table 4 is stronger than
$V^t_{\Lambda N}$ in a two-body sense, it is the effective
ranges that control the situation.  The scattering lengths are
essentially the same.  Because r$_0$($^1S_0$) is significantly
larger than r$_0$($^3S_1$), the potential $V^S_{\Lambda N}$ is
weaker than the potential $V^t_{\Lambda N}$ in a full
four-body calculation.

However, the nuclear core in $^4_{\Lambda}$H is not an
elementary object.  It is a $(J = \frac{1}{2}$,
$T = \frac{1}{2}$) composite of three nucleons.  Converting a
$T=0$ $\Lambda$ into a $T=1$ $\Sigma$ necessarily brings in
$T = \frac{3}{2}$ $^3$H$^{*}$ states.  These lie high in the
continuum (just as in the case of $^4$He mentioned
previously) {\it and} have more spatial structure than the
$^3$H ground state.  Therefore, the
$\Lambda - \Sigma$ conversion in the medium is expected to be
suppressed compared to its effect in the free
$\Lambda N - \Sigma N$ interaction.  Taking an extreme model
approximation
of dropping the parts of the $V_{\Lambda N - \Sigma N}$
interaction that couple through $T = \frac{3}{2}$ $^3$H$^*$
states, one can construct appropriately modified $\Lambda N$
effective interactions which yield {\cite{GL88}}, in the
same full four-body calculation:
$$
B(0^+) = 9.6 \; {\rm MeV}
$$
$$
B(1^+) = 8.2 \; {\rm MeV} .
$$
That is, one obtains a proper ordering of the states.  More
importantly, this demonstrates that $E_{\gamma}$ is
not a measure of the $\Lambda N$ spin-spin interaction.

How realistic are such separable potential calculations?
Monte Carlo variational calculations made with the Nijmegen
soft core potential have obtained $B(0^+) \simeq 1.5$ MeV,
while the $1^+$ state is unbound {\cite{JAC91}}.  The
$\Lambda - \Sigma$ conversion suppression is confirmed,
although it is too severe to match the data.  Our analysis
indicates that the $YN$ force modeling problem lies in the
sizeable strength in the $V_{\Lambda N - \Sigma N}$ tensor
potential in the spin-triplet channel.  Second order
tensor effects become very important when effectively
2 $\frac{1}{2}$ $YN$ pairs reside in the spin-triplet state.  Such
effects are well known in few-nucleon physics and play
a significant role also in hypernuclei.  The $\Lambda N - \Sigma N$
tensor coupling component provides so much
of the free interaction strength in the Nijmegen soft
core potential model that suppression of that coupling
leaves the spin-triplet dominated $1^+$ state unbound.  An
r-space approximation to J\"{u}lich \~A is needed in order
to test that model.

To reiterate, the $0^+ - 1^+$ energy difference does seem
to result from complex few-body dynamics and is not a
simple measure of the $\Lambda N$ spin-spin interaction.
It is possible to apply this result to understand
experiments involving heavier hypernuclei.  One can
prescribe a scenario based upon the separable potential
model results outlined here which may explain the failure
to find a $2^- \rightarrow 1^-$ $\gamma$ following the
$^{10}$B($K^-, \pi^-)^{10}_{\;\,\Lambda}$B$^*$ reaction
{\cite{chr90}}.  The prediction of such a transition is
based upon a shell model calculation in which the $A=4$
$0^+ - 1^+$ level splitting is used to define the
$\Lambda N$ spin-spin component of the mean field
$\Lambda$ potential {\cite{dal78,mil85}}.  In such a mean
field approach, the $0^+ - 1^+$ splitting in $A=4$ leads
to a prediction that the state ordering in
$^{10}_{\;\,\Lambda}$B is such that the ground state is $1^-$.
The $3^+$ g.s.\ of the target $^{10}$B ensures that the
$2^-$ state in $^{10}_{\;\,\Lambda}$B is produced.  Thus, a
$\gamma$ transition from the $2^-$ excited state to the
$1^-$ ground state should be seen.   Chrien {\it et al.}\
observed none between 0.1 and 0.9 MeV {\cite{chr90}}.
However, consider a cluster model in which
$^{10}_{\;\,\Lambda}$B is modeled as an
$\alpha$-$\alpha$-$\Lambda$-$p$  system.  The $\alpha$'s
are tightly bound.  The level structure of
$^{10}_{\;\,\Lambda}$B will likely be determined by the
spin dependence of the $\Lambda N$ interaction, even
though the valence proton lies in the p-shell while
the $\Lambda$ lies in the s-shell which reduces the
role of their direct spin-spin interaction and brings
into play spin-orbit and tensor forces.  That is, the
state ordering will likely correspond to that of the free
$YN$ interaction with no $\Lambda N - \Sigma N$ coupling
suppression.  (Recall that such an assumption in
the $A = 4$ system gave rise to a $1^+$ ground state.)
Therefore, one would anticipate that the ground state of
$^{10}_{\;\,\Lambda}$B is likely to be $2^-$ (not $1^-$),
and no $\gamma$ should have been seen in the experiment,
because the $1^-$ state is not produced in the reaction.
Clearly $^{10}_{\;\,\Lambda}$B is an important candidate for
further investigation, but it does appear that
$\Lambda N - \Sigma N$ coupling plays an important role
in the structure of $\Lambda$--hypernuclei.

\vspace{16pt}
\noindent {\bf 5. Hypernuclear Constraints}

\vspace{8pt}
In addition to the strong constraints that the
$^4_{\Lambda}$H--$^4_{\Lambda}$He isodoublet places upon
any model of the $YN$ interaction, $^3_{\Lambda}$H and
$^5_{\Lambda}$He also constrain various aspects of the
$YN$ interaction.  The different hypernuclei act as spin
filters, emphasizing different components of the $YN$
potential.

\noindent {\it 5.1. The Hypertriton}
\vspace{4pt}

Because the $YN$ interaction does not support a two-body bound
state {\cite{jjs62}}, as one can see from the scattering lengths
in Table 2, the hypertriton in the $S = -1$ sector serves the
function of the deuteron.  The fact that
$B_{\Lambda} \simeq 130 \pm 50$ keV
suggests that the long range properties of this
system should dominate.  However, the spin-singlet interaction
plays the dominant role in the hypertriton
$$
V_{\Lambda N} = \frac{3}{4} V^s_{\Lambda N} + \frac{1}{4}
V^t_{\Lambda N} \: ,
$$
so that $B(^3_{\Lambda}H)$ provides a constraint primarily upon
that potential component.

$\Lambda N - \Sigma N$ coupling has been known to play an
important role in $^3_{\Lambda}$H since the early calculations
of Schick {\cite{lhs68}}.  More recently, we have explored how
this translates into a three-body force (3BF) effect when the
$\Sigma N$ channel is formally eliminated {\cite{ira90}}.  In Fig.\ 1
are represented three types of diagrams that enter.  In (a) we
illustrate the usual iteration of the $\Lambda N$ potential
in $^3_{\Lambda}$H.  In (b) we see what is termed the dispersive
energy dependence of the $\Lambda N$ interaction when
the $\Sigma N$ channel is eliminated.  It is effectively
repulsive, because the energy denominator in the $\Sigma$
propagator becomes larger when the kinetic energy of the second
nucleon is added:
$$
V_{\Lambda N \rightarrow \Sigma N} \; \frac {1}{\Delta E} \;
V_{\Sigma N \rightarrow \Lambda N} .
$$
This effect is analogous to the reduction in strength of the
$NN$ tensor force when one embeds it in nuclear matter.  In
(c) we find the true 3BF that results.  The 3BF due to
$\Lambda N - \Sigma N$ coupling is actually attractive.  One is
led to ask whether the hypertriton would be bound were it not for
this attraction in a realistic potential calculation.

\begin{figure} [t]
\vspace*{8cm}
\begin{flushleft}
{\bf Fig.~1} Schematic representation of various interactions
among the $YN$ constituents of $^3_{\Lambda}$H:  a) a standard
contribution to the $\Lambda NN$ three-body equations, b) an
energy-dependent contribution to the $\Lambda N$ two-body $t$
matrix arising from $\Lambda N - \Sigma N$ coupling, and c) an
effective three-body force arising from $\Lambda N - \Sigma N$
coupling.
\end{flushleft}
\end{figure}

Miyazawa and Gl\"{o}ckle {\cite{miy93}} have recently completed
a model calculation of $^3_{\Lambda}$H using the J\"ulich OBE
model \~A {\cite{jul92}}.  They find the system unbound.  A 4\%
increase in the strength of $V^s_{YN}$ would produce binding
{\it but} would destroy the model \~A description of the $YN$
data.  This result is somewhat puzzling --- the $a,r_o$ values
for the model \~A imply significant attraction in a rank-one
separable potential $^3_{\Lambda}$H calculation:
$B_{\Lambda} \simeq 1$ MeV.  Why should the short range potential
properties play such a significant role in this weakly bound system?

\vspace{8pt}
\noindent {\it 5.2. The} $^5_{\Lambda}$He {\it Anomaly}
\vspace{4pt}

As discussed above, simple estimates for $B_{\Lambda}$ would
yield values about a factor of 2 larger than the data.  Model
calculations confirm this {\cite{her67,gal75,GGW72,GGW72a,bod88}}.
The anomalously small binding would appear to be due to two
sources:  1) a tensor force is less effective in binding a
few-body system than a central force with the same low energy
two-body properties, and 2) $\Lambda N - \Sigma N$ coupling is
suppressed because the $T=1$ $\Sigma$ must couple to $T=1$
even parity states high in the $\alpha$ spectrum.   Monte Carlo
variational calculations for the Nijmegen soft core model
support this analysis {\cite{JAC91}}.  These microscopic
calculations include explicit $\Sigma$ degrees of freedom.
The $NN$ force was represented by the Nijmegen '78 potential
{\cite{nij78}} and the Urbana model 7 three-nucleon interaction
{\cite{urb7}} was included to better represent the ground-state
properties of the three- and four-nucleon systems.  The $YN$
tensor force was shown to play a significant role; replacing
the YN spin-triplet interaction by the central force
spin-singlet intercation led to significant overbinding.
In contrast to the hypertriton, it is the spin-triplet
interaction that dominates
$$
V_{\Lambda N} = \frac{1}{4} V^s_{\Lambda N} + \frac{3}{4}
V^t_{\Lambda N} \: .
$$
Thus, $^5_{\Lambda}$He provides a significant constraint upon
that potential component.

\vspace{16pt}
\noindent {\bf 6. The} $S=-2$ {\bf Sector and the}
$^{\;\;\, 6}_{\Lambda \Lambda}$He {\bf Enigma}
\vspace{4pt}

The doubly-strange hypernuclei of the $S=-2$ sector provide
another handle on the importance of coupling to higher-lying
channels in the baryon-baryon interaction.
The single reported event {\cite{pro66}} is controversial
{\cite{dal89}}.  However $B_{\Lambda \Lambda} \simeq 10.6$
yields a value of
$$
<V_{\Lambda \Lambda}> = B_{\Lambda \Lambda} - 2 B_{\Lambda}
(^5_{\Lambda}{\rm He})
$$
$$
\simeq 4.4 \; {\rm MeV} \: ,
$$
which is consistent with the values of $4-5$ MeV extracted
from the other two $\Lambda \Lambda$ hypernuclei
{\cite{dan63,aok91,dov91}}.  Furthermore, Hartree-Fock
calculations {\cite{GGW69}} established that comparable
$V_{\Lambda \Lambda}$ and $V_{\Lambda N}$ interactions were
required to account for the binding energies of
$^{\;\;\, 6}_{\Lambda \Lambda}$He and $^4_{\Lambda}$He, respectively.
That is, one finds
$$
 <V_{\Lambda \Lambda}>_{A=6} \; \simeq \; <V_{\Lambda N}>_{A=4}
$$
could account for the binding energy of both systems.  The
$\Lambda N$ interaction is only weakly attractive, which implies
a similar property for the $\Lambda \Lambda$ force.
The preferred values of the scattering length and effective
range of the phenomenological $\Lambda \Lambda$ potential
required to reproduce the binding energies of the
$\Lambda \Lambda$ hypernuclei in the
analysis by Bodmer {\it et al.}\ {\cite{buc84}}
compared to those for the $\Lambda N$ system
listed in Table 2 lead to a similar conclusion, as does the
G-matrix analysis of Bando {\cite{ban82}} based upon the
Nijmegen model D.

However, just as $\Lambda N - \Sigma N$ coupling suppression
appears to play a significant role in the
$^4_{\Lambda}$He--$^4_{\Lambda}$H level ordering and in the
anomalously small binding of $^5_{\Lambda}$He, we hypothesize
that $\Lambda \Lambda - \Xi N$ coupling is likely to be
suppressed in $^{\;\;\, 6}_{\Lambda \Lambda}$He, so that only
$<V_{\Lambda \Lambda}>$ is sampled.  The $\Lambda \Lambda - \Xi N$
interaction lies in the same multiplet as the $nn$
interaction, which almost binds.  Thus, it is rather
surprising that $<V_{\Lambda \Lambda}>$, as extracted from
$^{\;\;\, 6}_{\Lambda \Lambda}$He, is weak by comparison.
Suppression of the $\Xi N$ conversion is a prime candidate to
explain this observation.  One anticipates such suppression
in $^{\;\;\, 6}_{\Lambda \Lambda}$He because conversion from
$\Lambda \Lambda \to \Xi N$ in a relative $^1$S$_0$ state
such that the resulting $\Xi N$ pair reside in the 1s shell
requires excitation of the $^4$He core.  Five nucleons
cannot exist in the 1s shell.

Therefore, we suggest that $^{\;\;\, 6}_{\Lambda \Lambda}$He
probes primarily the diagonal element $V_{\Lambda \Lambda}$
of the $\Lambda \Lambda - \Xi N$ interaction, while it would be
$^{\;\;\, 4}_{\Lambda \Lambda}$H (or $^{\;\;\, 5}_{\Lambda \Lambda}$He)
that would probe the full $\Lambda \Lambda - \Xi N$ force.
The $\Lambda \Lambda np$ system can easily couple to
$\Xi NNN$ in a manner that leaves all four baryons in the
1s shell.  Because $^3_{\Lambda}$H is bound,
$^{\;\;\, 4}_{\Lambda \Lambda}$H will also be bound.  If the full
$\Lambda \Lambda - \Xi N$ interaction strength is comparable
to that of the $nn$ interaction, then one could even
envision a $0^+$ particle-stable excited state as well
as a $1^+$ ground state for $^{\;\;\, 4}_{\Lambda \Lambda}$H.

\vspace{16pt}
\noindent {\bf 7. Summary}
\vspace{4pt}

In brief conclusion, we reiterate
the important physics differences between conventional,
nonstrange nuclear physics and hypernuclear physics.  The
few-baryon systems provide filters through which one can look at
particular aspects of the $YN$ interaction.  They provide specific
constraints on models of the interaction, and are particularly
sensitive to $\Lambda N - \Sigma N$ coupling and its induced
spin dependence.  Furthermore,
extrapolating our experience in the $S=-1$ sector to the $S=-2$
sector leads us to predict strong suppression of
$\Lambda \Lambda - \Xi N$ coupling in all but the lightest of
the $\Lambda \Lambda$ hypernuclei.

\vspace{12pt}
\noindent {\bf Acknowledgement}
\vspace{4pt}

The work of BFG and JAC was performed under the auspices of the
U.~S.~Department of Energy, while that of DRL was supported in
part by the U.\ S.\ Department of Energy under Grant No.\
DE-FG05-86ER40270, and that of IRA was supported by the Australian
Research Council.  BFG gratefully acknowledges a Research Award
for Senior Scientists by the Alexander von Humboldt Stiftung which
made possible his stay with the I.~K.~P.~Theorie group at the
Forschungszentrum J\"ulich.

\vspace{24pt}

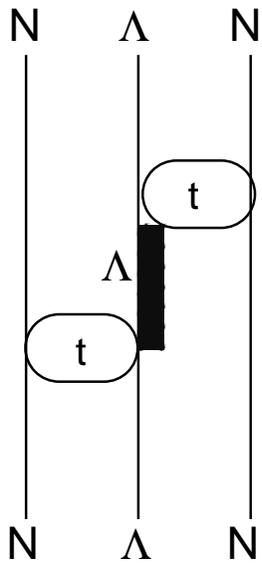
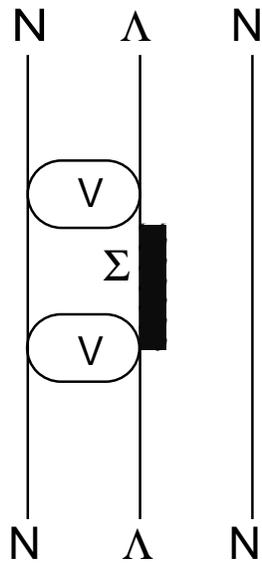
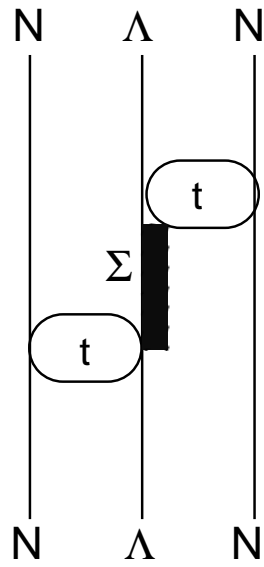
\end{document}